\documentclass[a4paper,11pt]{article}
\usepackage{amsmath,amsfonts,amssymb,amsthm,color,enumerate}
\usepackage{graphicx}
\usepackage{slashed}

\usepackage{braket}

\usepackage{a4wide}

\newcommand{\tx}{\text}
\newcommand{\J}{\tx{J}} 
\newcommand{\E}{\tx{E}} 
\newcommand{\MP}{M_\text{P}}

\newcommand{\Or}[1]{\ensuremath{\mathcal O\!\left(#1\right)}}

\newcommand{\R}{\tx{R}} 
\newcommand{\B}{\tx{B}} 

\newcommand{\G}{\mathcal G}
\newcommand{\M}{\mathcal M}

\newcommand{\red}[1]{{\color{black}#1}}

\newcommand{\magenta}[1]{{\color{black}#1}}

\newcommand{\paren}[1]{\left(#1\right)}
\newcommand{\fn}[1]{\!\left(#1\right)}
\newcommand{\sqbr}[1]{\left[#1\right]}
\newcommand{\br}[1]{\left\{#1\right\}}

\newcommand{\df}{\text{d}}
\newcommand{\ol}{\overline}
\newcommand{\al}[1]{\begin{align}#1\end{align}}
\newcommand{\als}[1]{\begin{align*}#1\end{align*}}

\newcommand{\mc}{\mathcal}
\newcommand{\bs}{\boldsymbol}

\newcommand{\nn}{\nonumber\\}

\usepackage{comment}

\DeclareMathOperator{\Tr}{Tr}
\DeclareMathOperator{\Det}{Det}
\newcommand{\pr}{\prime}

\title{
\vbox{
\baselineskip 14pt
\hfill \hbox{\normalsize OU-HET/906}\\
\hfill \hbox{\normalsize KEK-TH-1926}\\
\hfill \hbox{\normalsize MAD-TH-16-09}
}		
\vfill
Meaning of the field dependence of the renormalization scale in Higgs inflation
}

\author{
	Yuta Hamada,\thanks{\tt
		yhamada@wisc.edu}
		$^{1,2}$\
	Hikaru Kawai,\thanks{\tt
		hkawai@gauge.scphys.kyoto-u.ac.jp
		}
		$^3$\
	Yukari Nakanishi,\thanks{\tt
		nakanishi@het.phys.sci.osaka-u.ac.jp
		}
		$^4$\
	and
	Kin-ya Oda\thanks{\tt
		odakin@phys.sci.osaka-u.ac.jp
		}
		$^4$\bigskip\\
$^1$\it\normalsize
KEK Theory Center, IPNS, KEK, Tsukuba, Ibaraki 305-0801, Japan\\
$^2$\it\normalsize
Department of Physics, University of Wisconsin, Madison, WI 53706, USA\\
$^3$\it\normalsize
Department of Physics, Kyoto University, Kyoto 606-8502, Japan\\
$^4$\it\normalsize
Department of Physics, Osaka University, Osaka 560-0043, Japan\\
	\bigskip\\
	}
\date{\today}

\begin{document}

\maketitle

\begin{abstract}\noindent
We consider the prescription dependence of the Higgs effective potential under the presence of general nonminimal couplings.
We evaluate the fermion loop correction to the effective action in a simplified Higgs-Yukawa model whose path integral measure takes simple form either in the Jordan or Einstein frame.
The resultant effective action becomes identical in both cases when we properly take into account the quartically divergent term coming from the change of measure. 
Working in the counterterm formalism, we clarify that the difference between the prescriptions I and II comes from the counter term to cancel the logarithmic divergence.
This difference can be absorbed into the choice of tree-level potential from the infinitely many possibilities, including all the higher-dimensional terms.
We also present another mechanism to obtain a flat potential by freezing the running of the effective quartic coupling for large field values, using the nonminimal coupling in the gauge kinetic function.
\end{abstract}

\vfill
\newpage


\section{Introduction}
The Higgs inflation~\cite{Salopek:1988qh,Bezrukov:2007ep} is one of the closest to the best fit point in the tensor-to-scalar ratio vs spectral-index plane among various inflation models~\cite{Ade:2015lrj}.
The model requires rather a large nonminimal coupling $\xi\sim 10^{5\text{--}6}$ between the Higgs-squared $H^\dagger H$ and the Ricci scalar~$\mc R$.\footnote{
The earlier model~\cite{Salopek:1988qh,vanderBij:1993hx,CervantesCota:1995tz,vanderBij:1994bv}, without the Einstein-Hilbert action at the tree level, requires $\xi\sim 10^{34}$~\cite{Bezrukov:2007ep}.
In Ref.~\cite{Salopek:1988qh}, the authors have also studied the Higgs inflation model in Ref.~\cite{Bezrukov:2007ep} with essentially the same parameter $\xi\sim 10^4$ and $\lambda\sim \paren{\xi/10^5}^2\sim10^{-2}$.
See also Refs.~\cite{Lucchin:1985ip,Futamase:1987ua} for inflation with the nonminimal coupling, and Ref.~\cite{Ema:2016dny} for a possible issue with the large nonminimal coupling.
}
\red{On the other hand, the observed value of the Higgs mass $m_H=125.09\pm0.24\,\text{GeV}$~\cite{Aad:2015zhl} indicates that the Standard Model (SM) is at the \emph{criticality}, that is, the Higgs potential becomes small and nearly flat when the Higgs-field value is close to the Planck-scale; see e.g.\ Refs.~\cite{Hamada:2012bp,Hamada:2014wna}.

In the flat spacetime, the renormalized Higgs potential $V$ can be computed as a sum of the tree-level potential $V_\R$ and the loop correction $\Delta V_\R$, both of them being finite but depending on the renormalization scale $\mu$, in the counterterm formalism.
Since $V$ is independent of $\mu$, we may choose $\mu$ arbitrarily.
A convenient choice is $\mu\sim\varphi$, where $\varphi$ is the Higgs field value.
This choice minimizes $\Delta V_\R$, and then $V$ can be approximated by the tree-level potential: $V\simeq\left.V_\R\right|_{\mu=\varphi}$.

When we couple this system with gravity, in general, there arise corrections from the non-renormalizable couplings such as $\xi$. Under the presence of $\xi$, it has been said that there are two different ``prescriptions'' in which the renormalized Higgs potential is approximated by the tree-level one with~\cite{Bezrukov:2008ej,Bezrukov:2009db,Allison:2013uaa}}
\al{
\mu
	&\sim	\begin{cases}
				\displaystyle{\varphi\over\sqrt{1+\xi{\varphi^2\over \MP^2}}} & \tx{in prescription~I,}\\
				\varphi & \tx{in prescription~II,}
			\end{cases}
			\label{prescriptions given}
}
where $\MP =1/\sqrt{8\pi G}=2.4\times10^{18}\,\tx{GeV}$ is the reduced Planck scale.
The prescriptions I and II \red{are claimed to correspond to} a \red{$\varphi$}-independent ultraviolet (UV) cutoff in the Einstein and Jordan frames, respectively~\cite{Bezrukov:2008ej,Bezrukov:2009db}.
\red{Here the Jordan frame refers to the original action with a non-vanishing $\xi$, while the Einstein frame is the one without $\xi$, obtained by the field re-definition of the metric $g_{\mu\nu}$.
Note that if we introduce a UV cutoff as a $\varphi$-independent constant in either frame, then it becomes dependent on $\varphi$ in the other frame.
}
\red{If we accept Eq.~\eqref{prescriptions given} literally,}
the value of $\xi$ can be as low as of order 10 in the prescription~I~\cite{Hamada:2014iga,Bezrukov:2014bra,Hamada:2014wna} and $10^2$ in II~\cite{Hamada:2014iga,Hamada:2014wna}, under the SM criticality.
\red{The physical difference comes from the different large $\varphi$ limit of Eq.~\eqref{prescriptions given}.}

In this paper, we will revisit the relation between the UV cutoff and the renormalization scale.
\red{We clarify that the different choice of the cutoffs does not directly lead to the difference in Eq.~\eqref{prescriptions given}.}
We argue that, in the counterterm formalism, \red{it} may be regarded as the choice of the counter term to cancel the logarithmic divergence, and can be absorbed into the choice of the tree-level potential from infinitely many possibilities.

For that purpose, we consider the fermion loop correction to the effective action in a simplified Higgs-Yukawa model.
This toy model captures the essential features necessary to grasp what is going on in the realistic Higgs inflation model: 
As in the real world, we neglect the Higgs mass term which is much smaller than the one from quartic coupling $\lambda_4$ at the large field values under consideration;
the renormalization group (RG) running of $\lambda_4$ is governed by the loop of top-quark, which is represented by $\psi$.\footnote{
In reality, the loop of gauge bosons also contributes to the running of $\lambda_4$. However, the $\varphi$-dependent effective mass of the canonically normalized gauge boson, $g\varphi$, has the same $F_{\mc R}\fn{\varphi}$ dependence in the Einstein frame, $g\varphi/\sqrt{F_{\mc R}\fn{\varphi}}$, as the effective mass of fermion $y\varphi$ which becomes $y\varphi/\sqrt{F_{\mc R}\fn{\varphi}}$; see e.g.\ Ref.~\cite{Hamada:2014wna}.
Therefore the arguments for frame independence and for prescription dependence should apply without modification after we include gauge boson loops.
}

This paper is organized as follows.
In Sec.~\ref{loop correction section}, we obtain the one-loop effective action in the Higgs-Yukawa model both in the Jordan and Einstein frames. We show that the effective action is independent of the frame if we properly take into account the change of the path integral measure. This change of measure affects only the quartically divergent term, and has nothing to do with the difference between the prescriptions I and II that is related to the logarithmic divergence.
In Sec.~\ref{prescriptions}, in the counterterm formalism, we show that the difference between prescriptions I and II comes from the choice of the counter term to cancel the logarithmic divergence.
We point out that this difference can be absorbed into the choice of the tree-level potential, including higher dimensional terms, from the infinite possibilities.
In Sec.~\ref{gauge kinetic}, 
we present a mechanism that uses the gauge kinetic function to stop the running of effective quartic coupling for large $\varphi$ as in the prescription I in Eq.~\eqref{prescriptions given}, which helps to further flatten the Higgs potential at high scales.
In the last section, we summarize our result.

\section{Quantum correction from fermion loop}\label{loop correction section}

\subsection{Frames at classical level}
We first review the transformation from Jordan to Einstein frames at the classical level. Our starting action in Jordan frame is
\al{
S
	&=	\int \df^4 x\sqrt{-g}\Bigg[
		\frac{\MP^2}{2}F_{\mc R}\fn{\varphi}\mc R
		-\frac{1}{2}F_\Phi\fn{\varphi}g^{\mu\nu}\,\partial_{\mu}\varphi\,\partial_{\nu}\varphi
	-V\fn{\varphi}\nn
	&\phantom{=	\int \df^4 x\sqrt{-g}\Bigg[}
		-F_\Psi\fn{\varphi}\ol{\psi}\gamma^{\mu}D_{\mu}\psi
	-F_\tx{Y}\fn{\varphi}y\varphi\ol{\psi}\psi
	\Bigg],
	\label{Jordan frame action}
}
where
$y$ is the Yukawa coupling\footnote{
This $y$ is related to the SM top Yukawa coupling $y_t$ by $y=y_t/\sqrt{2}$.
}
and
$D_\mu=\partial_\mu+\Omega_\mu$ is the general covariant derivative on spinor, with $\Omega_\mu$ being the spin-connection.
We assume that we may take a weak-field limit $\varphi\to0$ so that we can expand the action around $\varphi=0$. That is,
\al{
F_X\fn{\varphi}
	&=	1+\xi_X{\varphi^2\over M^2}+\cdots
		\label{F expanded}
}
for $X=\mc R$, $\Phi$, $\Psi$ and Y, where $\xi_X$ is the first-order nonminimal coupling and $M$ is the typical scale of UV theory, such as the string scale.\footnote{\label{relation to ordinary xi}
The nonminimal coupling $\xi$ between Higgs and Ricci scalar in the ordinary notation reads $\xi=\xi_{\mc R}{\MP^2/M^2}$~\cite{Bezrukov:2007ep}. See also Ref.~\cite{Broy:2016rfg} for more arbitrary extension with large nonminimal couplings.
}
We have also assumed for simplicity that the action is invariant under a chiral $Z_2$ symmetry
\al{
\varphi	&\to	-\varphi,	&
\psi	&\to	\gamma_5\psi.
}
In general, the potential contains all the higher dimensional terms:
\al{
V\fn{\varphi}
	&=	\sum_\tx{$n$: even, $n\geq0$}{\lambda_n}{\varphi^n\over M^{n-4}}.
		\label{Jordan frame V}
}

By the field redefinition
\al{
g^\E_{\mu\nu}=F_{\mc R}\fn{\varphi}g_{\mu\nu},
	\label{metric redefinition}
}
we obtain the Einstein-frame action\footnote{
In Ref.~\cite{Hamada:2014wna}, the factor in front of the scalar kinetic term has a typo and should read $\sqbr{{B\over A}+{3\over2}{M_P^2A^{\pr2}\over A^2}}$.
}
\al{
S
	&=	\int \df^4 x\sqrt{-g_\E}\Bigg[
		\frac{\MP^2}{2}\mc R_\E
		-{1\over2}\paren{{F_\Phi\fn{\varphi}\over F_{\mc R}\fn{\varphi}}
			+{3\over2}\paren{\MP F_{\mc R}'\fn{\varphi}\over F_{\mc R}\fn{\varphi}}^2}g_\E^{\mu\nu}\,\partial_{\mu}\varphi\,\partial_{\nu}\varphi
		-{V\fn{\varphi}\over \paren{F_{\mc R}\fn{\varphi}}^2}\nn
	&\phantom{=	\int \df^4 x\sqrt{-g_\E}\Bigg[}
		-{F_\Psi\fn{\varphi}\over\paren{F_{\mc R}\fn{\varphi}}^{3/2}}\ol{\psi}\gamma_\E^{\mu}D_{\mu}^\E\psi
		-{F_\tx{Y}\fn{\varphi}\over \paren{F_{\mc R}\fn{\varphi}}^2}y\varphi\ol{\psi}\psi
		\Bigg].
		\label{Einstein frame action}
}
Here and hereafter, we put either sub- or superscript ``E'' and ``J'' on quantities in the Einstein and Jordan frames, respectively, when it is preferable; the ones without such sub- or superscript are given in the Jordan frame unless otherwise stated.

The original Higgs inflation~\cite{Bezrukov:2007ep} assumes that the potential~\eqref{Jordan frame V} can be approximated by
\al{
V\fn{\varphi}
	&=	{\lambda_4}\varphi^4,
		\label{phi4 potential}
}
namely $\lambda_n\ll1$ for $n\neq4$, at around the scale $\varphi\sim M\lesssim\MP$.\footnote{\label{footnote on lambda}
For the Higgs inflation under SM criticality~\cite{Hamada:2014iga,Hamada:2014wna}, we further assume the flatness $\lambda_4\ll1$ at high scales. In terms of the quartic coupling $\lambda$ in Refs.~\cite{Hamada:2014iga,Hamada:2014wna}, the quartic coupling in this paper is written as $\lambda_4=\lambda/4$.
}
That is, one assumes that all the higher order terms are small at $\varphi\sim M$ so that $V\ll M^4\lesssim\MP^4$.
Combined with the assumption $F_{\mc R}=1+\xi{\varphi^2/\MP^2}$, the potential in Eq.~\eqref{Einstein frame action},
\al{
U	:=	{V\over F_{\mc R}^2},
	\label{Einstein-frame potential}
}
becomes constant in the large $\varphi$ limit:
\al{
U
	&\to	{\lambda_4\over\xi^2}\MP^4,
		\label{Utree}
}
leading to the inflation.
\red{The field $\varphi$ is not canonically normalized in the Einstein frame, and the change of equation of motion must be taken into account; see Appendix~\ref{mechanism}.}
Note that it is important to terminate the expansion of $V$ and $F_{\mc R}$ at $\varphi^4$ and $\varphi^2$, respectively, in order to obtain this constant potential.\footnote{\red{
The other option is to terminate them at the $2n$th and $n$th orders, respectively; see the last point in Appendix~\ref{various}.
}}


\subsection{Frame dependence of UV cutoffs}\label{frame dependent cutoffs}
Now we take into account the quantum corrections.
In this paper, we evaluate the one-loop correction to the effective action from the fermion loop, leaving those from the graviton and $\phi$ loops.\footnote{
The correction from $\phi$ loop is proportional to $\lambda_n$, which are assumed to be small here;
see Refs.~\cite{Hamada:2012bp,Hamada:2013mya,Hamada:2015ria} for arguments in support of the smallness of $\lambda_n$ at high scales.
}
We compute the corrections to $\lambda_n$ only.
That is, we neglect all the corrections to other couplings $y$, $\xi_X$, etc., and hence do not distinguish the bare and renormalized couplings for them.

In a given frame, short-distance cutoff $\ell$ is given by
\al{
\ell^2	=	g_{\mu\nu}\,\Delta x^\mu\,\Delta x^\nu.
}
Then the metric redefinition~\eqref{metric redefinition} relates the cutoff lengths in two frames by
\al{
\ell_\J^2
	&=	g_{\mu\nu}^\J\,\Delta x^\mu\,\Delta x^\nu
	=	{g_{\mu\nu}^\E\over F_{\mc R}\fn{\varphi}}\Delta x^\mu\,\Delta x^\nu
	=	{\ell_\E^2\over F_{\mc R}\fn{\varphi}}.
}
That is, the UV cutoff scales are related by~\cite{Bezrukov:2008ej}
\al{
\Lambda_\E^2
	&=	{\Lambda_\J^2\over F_{\mc R}\fn{\varphi}}.
		\label{two cutoffs}
}
As pointed out in Refs.~\cite{Bezrukov:2008ej}, we may choose either $\Lambda_\J$ or $\Lambda_\E$ to be independent of $\varphi$, but not both.\footnote{
We note that such a field-dependent cutoff itself does not lead to any logical inconsistency. For example, a position-dependent momentum cutoff follows from the Pauli-Villars regularization in warped space even if we start from a position-independent bulk mass for the regulator~\cite{Pomarol:2000hp}.
}
In the prescriptions I and II in the original sense~\cite{Bezrukov:2008ej}, we set $\Lambda_\E$ and $\Lambda_\J$ to be a constant, respectively.

\subsection{Frame independence of effective action up to quartic divergence}
In general, the effective action should not depend on the choice of frame if we properly take into account the change of the path integral measure as well as that of the cutoff~\eqref{two cutoffs}.
We demonstrate it at the one-loop level under the simplifying assumption given above.

The one-loop effective action induced by the fermion loop in the Jordan frame is given by
\al{
e^{i\Delta S^\J_\tx{eff}}
	&:=	\int \mc D_{g_\J}\psi\ \mc D_{g_\J}\ol\psi\ \exp\sqbr{i\int\df^4x\sqrt{-g_\J}\,\ol\psi\paren{-F_\Psi\slashed D_{g_\J}-F_\tx{Y}y\varphi}\psi}.
					\label{Jordan frame effective action}
}
There is no unique definition of the path integral measure $\mc D_{g_\J}\psi$.
Here we take a simple measure that is induced from the following distance in the functional space
\al{
\left\lVert\delta\psi\right\rVert_{g_\J}^2=\int\df^4x\sqrt{-g_\J}\,\delta\ol\psi\,\delta\psi,
	\label{functional measure}
}
which is invariant under the diffeomorphism.

Let us rewrite Eq.~\eqref{Jordan frame effective action} into the path integral in Einstein frame.
Because we have
\al{
\left\lVert\delta\psi\right\rVert_{g_\J}^2
	&=	\int\df^4x\sqrt{-{g_\J}}\,\delta\ol\psi\,\delta\psi\nn
	&=	\int\df^4x\sqrt{-g_\E}\,F_{\mc R}^{-2}\,\delta\ol\psi\,\delta\psi,
		\label{simple distance in Jordan}
}
the functional measure satisfies
\al{
\mc D_{g_\J}\psi\ \mc D_{g_\J}\ol\psi
	&=	\mc D_{g_\E}\psi\ \mc D_{g_\E}\ol\psi\ \paren{\prod_x F_{\mc R}^{-2}}^{-4}\nn
	&=	\mc D_{g_\E}\psi\ \mc D_{g_\E}\ol\psi\ \exp\sqbr{-4\,\underset{g_\E,\Lambda_\E}{\Tr}\ln F_{\mc R}^{-2}},
		\label{conformal tf J to E}
}
where the measure $\mc D_{g_\E}\psi$ is the one induced from the following distance in the functional space
\al{
\left\lVert\delta\psi\right\rVert_{g_\E}^2=\int\df^4x\sqrt{-g_\E}\,\delta\ol\psi\,\delta\psi,
	\label{distance in E}
}
and $\Tr_{g_\E, \Lambda_\E}$ indicates that the functional trace depends both on the metric $g_\E$ and the cutoff~$\Lambda_\E$; 
the more explicit form will be presented below.\footnote{
The extra minus sign of $-4$ in Eq.~\eqref{conformal tf J to E} is from the Jacobian for fermionic variables.
}

When we rewrite the functional measure $\mc D_{g_\J}\psi\ \mc D_{g_\J}\ol\psi$ in terms of $\mc D_{g_\E}\psi\ \mc D_{g_\E}\ol\psi$, there has appeared the extra contribution to the effective action:
\al{
\exp\sqbr{-4\,\underset{g_\E,\Lambda_\E}{\Tr}\,\ln F_{\mc R}^{-2}}.
	\label{extra contribution}
}
As we will see below, this factor contains quartic divergence, and is absorbed into the renormalized couplings including the coefficients of higher dimensional terms.
Finally, Eq.~\eqref{Jordan frame effective action} becomes
\al{
e^{i\Delta S^\J_\tx{eff}}
	&=	\exp\sqbr{-4\,\underset{g_\E,\Lambda_\E}{\Tr}\,\ln F_{\mc R}^{-2}}
		\int{\mc D_{g_\E}\psi}\ {\mc D_{g_\E}\ol\psi}\ 
		\exp\sqbr{i\int\df^4x\sqrt{-g_\E}\,\ol\psi\paren{-{F_\Psi\over F_{\mc R}^{3/2}}\slashed D_{g_\E}-{F_\tx{Y}\over F_{\mc R}^2}y\varphi}\psi}\nn
	&=:	\exp\sqbr{-4\,\underset{g_\E,\Lambda_\E}{\Tr}\,\ln F_{\mc R}^{-2}}
		e^{i\Delta S^\E_\tx{eff}},
			\label{Einstein frame effective action}
}
where we have defined the Einstein-frame effective action $\Delta S_\tx{eff}^\E$, which is obtained from the path integral measure~$\mc D_{g_\E}\psi$.\footnote{
The same argument applies if we start from a different theory defined with another measure induced from the distance~\eqref{distance in E} instead of Eq.~\eqref{simple distance in Jordan}. Then the Jordan-frame effective action will receive extra contribution from the change of measure,
$\exp\sqbr{4\,\underset{g_\J,\Lambda_\J}{\Tr}\,\ln F_{\mc R}^{-2}}$,
which again will make the difference only in the renormalization conditions.
}
In this section hereafter, we will put the superscript $\J$ ($\E$) for the effective potential of the theory that is defined by using the measure $\mc D_{g_\J}\psi$ ($\mc D_{g_\E}\psi$), as well as the effective action.

\subsection{Explicit computations}
Now we verify the equality~\eqref{Einstein frame effective action} through more explicit computations under the assumption that $\varphi$ and $g_{\mu\nu}$ are slowly varying backgrounds so that they may be treated as constants in the computation of the effective action.

As a preparation, let us first compute the extra factor~\eqref{extra contribution} coming from the change of measure:
\al{
\exp\sqbr{-4\,\underset{g_\E,\Lambda_\E}{\Tr}\,\ln F_{\mc R}^{-2}}
	&=	\exp\sqbr{-4i\int\df^4x\sqrt{-g_\E}\Braket{x|\ln F_{\mc R}^{-2}|x}_{\Lambda_\E}}\nn
	&=	\exp\sqbr{i\int\df^4x\sqrt{-g_\E}\paren{-{\Lambda_\E^4\over8\pi^2}\ln F_{\mc R}^{-2}}},
		\label{extra factor}
}
where we have used
\al{
\Braket{x|x}_{\Lambda_\E}
	&=	\int^{\Lambda_\E}{\df^4p\over\paren{2\pi}^4}
	=	\int_0^{\Lambda_\E}{2\pi^2p^3\df p\over\paren{2\pi}^4}
	=	{\Lambda_\E^4\over32\pi^2}.
}
Here and hereafter, the momentum integral is taken in the Euclidean space.

The effective action~\eqref{Jordan frame effective action} reads
\al{
e^{i\Delta S^\J_\tx{eff}}
	&=
		\underset{g_\J,\Lambda_\J}{\Det}\fn{-{F_\Psi}\slashed D_{g_\J}-F_\tx{Y}y\varphi\over\mu_0}\nn
	&=	
		\exp\sqbr{\underset{g_\J,\Lambda_\J}{\Tr}\,\ln\fn{-F_\Psi\slashed D_{g_\J}-F_\tx{Y}y\varphi\over\mu_0}}\nn
	&=
		\exp\sqbr{4i\int\df^4x\sqrt{-g_\J}\int^{\Lambda_\J}{\df^4p\over\paren{2\pi}^4}\,{
			{1\over2}\ln\fn{F_\Psi^2p^2+F_\tx{Y}^2\paren{y\varphi}^2\over\mu_0^2}}
			}\nn
	&=: \exp\sqbr{i\int\df^4x\sqrt{-g_\J}\paren{-\Delta V^\J_\tx{eff}}},
		\label{effective action being computed}
}
where $\mu_0$ is an arbitrary reference scale and we have defined the correction to the Jordan-frame potential $\Delta V_\tx{eff}^\J$.
It may be computed as
\al{
\Delta V^\J_\tx{eff}
	&=	-2\int^{\Lambda_\J}{\df^4p\over\paren{2\pi}^4}\,
			\ln\fn{F_\Psi^2p^2+F_\tx{Y}^2\paren{y\varphi}^2\over\mu_0^2}\nn
	&=	-{1\over16\pi^2}\br{
			\Lambda_\J^4\sqbr{\ln\fn{F_\Psi^2{\Lambda_\J^2+\M_\J^2\over\mu_0^2}}-{1\over2}}
			+\Lambda_\J^2\M_\J^2
			+\M_\J^4\ln\fn{\M_\J^2\over\Lambda_\J^2+\M_\J^2}
			},
			\label{Jordan frame effective potential}
}
where
\al{
\M_\J\fn{\varphi}
	:=	y\varphi\,{F_\tx{Y}\fn{\varphi}\over F_\Psi\fn{\varphi}}
	\label{effective mass}
}
is the field-dependent mass for canonically normalized fermion in the Jordan frame.\footnote{
For more realistic top quark loop, $\Delta V_\tx{eff}$ is multiplied by the color degrees of freedom $N=3$.
}
We may rewrite the effective action~\eqref{effective action being computed} with the Einstein-frame metric:
\al{
e^{i\Delta S_\tx{eff}^\J}
	&=
		\exp\sqbr{i\int\df^4x\sqrt{-g_\E}\paren{-{\Delta V^\J_\tx{eff}\over F_{\mc R}^2}}}
	=:	\exp\sqbr{i\int\df^4x\sqrt{-g_\E}\paren{-\Delta U^\J_\tx{eff}}},
		\label{U vs V in J}
}
where we have defined the correction to the potential~\eqref{Einstein-frame potential}. That is,
\al{
\Delta U^\J_\tx{eff}
	=	{\Delta V^\J_\tx{eff}\over F_{\mc R}^2}
	&=	-{1\over16\pi^2F_{\mc R}^2}\br{
			\Lambda_\J^4\sqbr{\ln\fn{F_\Psi^2{\Lambda_\J^2+\M_\J^2\over\mu_0^2}}-{1\over2}}
			+\Lambda_\J^2\M_\J^2
			+\M_\J^4\ln\fn{\M_\J^2\over\Lambda_\J^2+\M_\J^2}
			}.
			\label{U in J}
}

To summarize, we have started from the measure~\eqref{functional measure}, and  computed the one-loop correction~\eqref{Jordan frame effective potential}.
\red{One may worry that the change of path measure~\eqref{conformal tf J to E} might introduce a trace anomaly in addition to Eq.~\eqref{extra factor}. However, it is taken into account as a form of the logarithmic UV cutoff dependence in Eq.~\eqref{Jordan frame effective potential}. Indeed, the constant shift of $\ln\Lambda_\J$ correctly reproduces the trace anomaly, as can be seen from Eq.~\eqref{Jordan frame V expanded} with Eq.~\eqref{effective mass}, compared to Eq.~\eqref{running of lambda}.}

We may instead perform the field redefinition~\eqref{metric redefinition} to the Einstein frame first, and compute the Einstein-frame effective action in the right-hand side of Eq.~\eqref{Einstein frame effective action}:
\al{
e^{i\Delta S^\E_\tx{eff}}
	&=	\exp\sqbr{\underset{g_\E,\Lambda_\E}{\Tr}\,\ln\paren{-{F_\Psi\over F_{\mc R}^{3/2}}\slashed D_{g_\E}-{F_\tx{Y}\over F_{\mc R}^2}y\varphi\over\mu_0}}\nn
	&=	\exp\sqbr{4i\int\df^4x\sqrt{-g_\E}\,\int^{\Lambda_\E}{\df^4p\over\paren{2\pi}^4}\,
			{1\over2}\ln\fn{{F_\Psi^2\over F_{\mc R}^3}p^2+{F_\tx{Y}^2\over F_{\mc R}^4}\paren{y\varphi}^2\over\mu_0^2}}\nn
	&=:	\exp\sqbr{i\int\df^4x\sqrt{-g_\E}\paren{-\Delta U^\E_\tx{eff}}},
		\label{Einstein frame effective action}
}
where $\Delta U_\tx{eff}^\E$ is the fermion loop correction to the potential~\eqref{Einstein-frame potential} that is obtained with the measure $\mc D_{g_\E}\psi$:
\al{
\Delta U^\E_\tx{eff}
	&=	-2\int_0^{\Lambda_\E}{\df^4p\over\paren{2\pi}^4}\sqbr{
			\ln\fn{{F_\Psi^2\over F_{\mc R}^3}p^2+{F_\tx{Y}^2\over F_{\mc R}^4}y^2\varphi^2\over\mu_0^2}
			}\nn
	&=	
		-{1\over16\pi^2}\br{
			\Lambda_\E^4\sqbr{\ln\fn{{F_\Psi^2\over F_{\mc R}^3}{\Lambda_\E^2+{\M_\J^2\over F_{\mc R}}\over\mu_0^2}}-{1\over2}}
			+\Lambda_\E^2{\M_\J^2\over F_{\mc R}}
			+{\M_\J^4\over F_{\mc R}^2}\ln\fn{{\M_\J^2\over F_{\mc R}}\over\Lambda_\E^2+{\M_\J^2\over F_{\mc R}}}
			}.
			\label{Einstein frame effective potential with Lambda E}
}

With the identification of UV cutoff scales~\eqref{two cutoffs}, we may rewrite
\al{
\Delta U^\E_\tx{eff}
	&=	-{1\over16\pi^2F_{\mc R}^2}\br{
			{\Lambda_\J^4}\sqbr{\ln\fn{{F_\Psi^2\over F_{\mc R}^4}{\Lambda_\J^2+\M_\J^2\over\mu_0^2}}-{1\over2}}
			+{\Lambda_\J^2}{\M_\J^2}
			+{\M_\J^4}\ln\fn{{\M_\J^2}\over{\Lambda_\J^2}+{\M_\J^2}}
			}.
			\label{Einstein frame effective potential}
}
We see that
\al{
\Delta U^\E_\tx{eff}
	&=	\Delta U^\J_\tx{eff}-{1\over16\pi^2}\Lambda_\E^4\ln F_{\mc R}^{-4}.
		\label{relation of two effective potentials}
}
Using Eq.~\eqref{extra factor}, we see that Eq.~\eqref{relation of two effective potentials} is equivalent to Eq.~\eqref{Einstein frame effective potential}.
We note that the difference in~\eqref{relation of two effective potentials} is quartically divergent, which will be subtracted by the renormalization.
In particular, this difference does not change the running of couplings, as we will see.

To summarize, once we fix the path integral measure, say, to be $\mc D_{g_\J}\psi$, we obtain the same result~$\Delta U^\J_\tx{eff}$, no matter in which frame we compute it:
When we compute it in the Jordan frame, we obtain
\al{
e^{i\Delta S_\tx{eff}^\J}
	&=	e^{i\int\df^4x\sqrt{-g_\J}\paren{-\Delta V_\tx{eff}^\J}}\nn
	&=	e^{i\int\df^4x\sqrt{-g_\E}\paren{-\Delta U_\tx{eff}^\J}},
}
while when we compute it in the Einstein frame,
\al{
e^{i\Delta S_\tx{eff}^\J}
	&=	e^{i\int\df^4x\sqrt{-g_\E}\paren{-{\Lambda_\E^4\over8\pi^2}\ln F_{\mc R}^{-2}}}
		e^{i\int\df^4x\sqrt{-g_\E}\paren{-\paren{\Delta U_\tx{eff}^\J-{\Lambda_\E^4\over16\pi^2}\ln F_{\mc R}^{-4}}}}\nn
	&=	e^{i\int\df^4x\sqrt{-g_\E}\paren{-\Delta U_\tx{eff}^\J}}.
}
We have explicitly checked that these two agree.


The frame independence of the effective potential has been verified in various ways:
In Refs.~\cite{George:2013iia,George:2015nza}, the authors have obtained one-loop RG equations for the tree level action in both the Jordan and Einstein frames, and have found the agreement between both results; see also appendix of Ref.~\cite{Kannike:2015apa}.
In Ref.~\cite{Postma:2014vaa}, the authors checked that both the tree-level actions are equivalent when written in terms of dimensionless variables, as it should be.
In Ref.~\cite{Kamenshchik:2014waa}, the authors have computed the one-loop divergent part of the effective potential in both frames, and have shown that both coincide at on-shell.
In~Refs.~\cite{Makino:1991sg,Domenech:2015qoa,Domenech:2016yxd}, the authors have discussed frame independence of physical observables.

\section{Prescriptions I and II}\label{prescriptions}
We now discuss the meaning of the prescriptions I and II.
We first clarify how the difference of the prescriptions in Eq.~\eqref{prescriptions given} arises in the ordinary context. Then we will show, in the counterterm formalism, that this difference can be absorbed into the choice from infinitely many possibilities of the coefficients of higher dimensional terms in the tree-level potential.

We consider the cutoff theory containing infinite number of higher dimensional terms.
We tune the infinite number of bare couplings in the large cutoff limit $\Lambda_\J,\Lambda_\E\to\infty$ such that the renormalized effective potential becomes a function of $\varphi/M$, where $M$ is the physical mass scale; see e.g.\ Ref.~\cite{Gomis:1995jp}.
We work in the counterterm formalism so that $\M_\J$ and $F_\Psi$ are treated as finite renormalized quantities.
To be concrete, we consider the theory defined by the path integral measure $\mc D_{g_\J}\psi$, and we omit the superscript $\J$ from the potentials $V^\J$, $\Delta V_\tx{eff}^\J$, \dots, etc.\ hereafter.\footnote{
Exactly the same argument applies if we consider the theory defined by the measure $\mc D_{g_\E}\psi$.
}

\subsection{Prescription II in ordinary context}
We clarify how the prescription II in Eq.~\eqref{prescriptions given} appears in the ordinary context.
The contribution from fermion to the effective potential~\eqref{Jordan frame effective potential} contains the quartic, quadratic, and logarithmic divergences:
\al{
\Delta V_\tx{eff}
	&=	-{1\over16\pi^2}\br{
			\Lambda_\J^4\sqbr{\ln\fn{F_\Psi^2{\Lambda_\J^2\over\mu_0^2}}-{1\over2}}
			+2\Lambda_\J^2\M_\J^2
			+\M_\J^4\paren{\ln{\M_\J^2\over\Lambda_\J^2}-{1\over2}}
			}
		+\Or{\Lambda_\J^{-2}}.
		\label{Jordan frame V expanded}
}
Then in the full effective potential
\al{
V=V_\B+\Delta V_\tx{eff},
	\label{full effective potential}
}
we cancel the divergences in Eq.~\eqref{Jordan frame V expanded} by the bare couplings $\lambda_{n\B}$ in the bare potential 
\al{
V_\B:=\sum_n\lambda_{n\B}{\varphi^n\over M^{n-4}}.
}
The quartic and quadratic divergences in Eq.~\eqref{Jordan frame V expanded} can be simply subtracted by the counter term
\al{
V^\tx{c.t.}_\tx{power}
	&=	{1\over16\pi^2}\br{
			\Lambda_\J^4\sqbr{\ln\fn{F_\Psi^2{\Lambda_\J^2\over\mu_0^2}}-{1\over2}}
			+2\Lambda_\J^2\M_\J^2}.
				\label{power ct}
}

However, we need a special care in subtracting the logarithmic divergence in Eq.~\eqref{Jordan frame V expanded} because the counter term should be analytic around $\varphi=0$:
A counter term having $\ln\M_\J=\ln\varphi+\cdots$ breaks the analyticity around $\varphi=0$.
In particular, the $(n+1)$th derivative of the term $\varphi^n\ln\varphi$ is singular, and the $(n+1)$-point function becomes ill-defined in the weak field limit $\varphi\to0$ if the bare action has such a term.\footnote{
In contrast, the singular behavior of $(n+1)$th derivative of the effective action represents the infrared singularity of the $(n+1)$-point scattering of massless scalar, which is cured by taking into account the Higgs mass in reality and/or by concentrating on the infrared-safe physical quantities.
}
Because we employ the analytic tree-level potential, the counter term should then be analytic too.

A natural choice of the counter term that is analytic around $\varphi=0$ would be
\al{
V^\tx{c.t.II}_\tx{log}
	&=	{\M_\J^4\over16\pi^2}
			\ln{\mu^2\over\Lambda_\J^2},
				\label{log ct}
}
where $\mu$ is the renormalization scale.
The resultant bare potential is
\al{
V_\B^\tx{II}
	&=	V_\R^\tx{\red{II}}+V^\tx{c.t.}_\tx{power}+V^\tx{c.t.II}_\tx{log}\nn
	&=	V_\R^\tx{\red{II}}
		+{1\over16\pi^2}\br{
			\Lambda_\J^4\sqbr{\ln\fn{F_\Psi^2{\Lambda_\J^2\over\mu_0^2}}-{1\over2}}
			+2\Lambda_\J^2\M_\J^2
			+\M_\J^4\ln{\mu^2\over\Lambda_\J^2}
			},
			\label{minimal subtraction}
}
where $V_\R^\tx{\red{II}}$ is the tree-level potential in the counterterm formalism.
%
We note that $V_\R^\tx{\red{II}}$ is $\mu$-dependent:
\al{
V_\R^\tx{\red{II}}\fn{\varphi,\mu}
	&=	\sum_n\lambda_{n\R}\fn{\mu}{\varphi^n\over M^{n-4}},
		\label{tree level potential}
}
and the $\mu$-independence of $V_\B$ determines the running of $\lambda_{n\R}\fn{\mu}$ via Eq.~\eqref{minimal subtraction}. In particular, because $\M_\J=y\varphi+\mc O\fn{\varphi^3}$, we obtain the ordinary running of the quartic coupling:
\al{
{\df\lambda_{4\R}\fn{\mu}\over\df\ln\mu}
	&=	-{y^4\over8\pi^2}.
		\label{running of lambda}
}

Substituting the bare potential~\eqref{minimal subtraction} into $V_\B$ in Eq.~\eqref{full effective potential}, we obtain
\al{
V\fn{\varphi}
	&=	V_\B^\tx{II}+\Delta V_\tx{eff}\nn
	&=	V_\R^\tx{\red{II}}\fn{\varphi,\mu}+\red{\Delta V^\tx{II}_\R\fn{\varphi,\mu},}
		\label{running potential in J}
}
\red{where}
\al{
\red{\Delta V_\R^\tx{II}\fn{\varphi,\mu}}
	:=	-{\sqbr{\M_\J\fn{\varphi}}^4\over16\pi^2}\paren{\ln{\sqbr{\M_\J\fn{\varphi}}^2\over\mu^2}-{1\over2}}
		\label{finite correction II}
}
\red{is the one-loop correction in the counterterm formalism in the prescription II.
Now both $V_\R^\tx{II}$ and $\Delta V_\R^\tx{II}$ are finite.}

\red{When we want to minimize the correction~\eqref{finite correction II},} we may choose the renormalization scale\footnote{
The constant $-1/2$ is scheme-dependent and does not affect our argument here.
}
\al{
\mu\sim\M_\J.
	\label{mu in II}
}
This result reproduces the prescription II in the sense of Eq.~\eqref{prescriptions given}, namely $\mu\sim\varphi$ for $F_\Psi=F_\tx{Y}=1$.

\subsection{Prescription I in ordinary context}
We clarify how the prescription I in Eq.~\eqref{prescriptions given} appears in the ordinary context.
We can rewrite Eq.~\eqref{Jordan frame V expanded} by using Eq.~\eqref{two cutoffs}:\footnote{
This may also be verified by substituting Eq.~\eqref{Einstein frame effective potential with Lambda E} into Eq.~\eqref{relation of two effective potentials}, dividing both-hand sides by $F_{\mc R}^2$, and expanding it in terms of $\Lambda_\E$.
}
\al{
\Delta V_\tx{eff}
	&=	-{F_{\mc R}^2\over16\pi^2}\br{
			\Lambda_\E^4\sqbr{\ln\fn{F_\Psi^2{F_{\mc R}\Lambda_\E^2\over\mu_0^2}}-{1\over2}}
			+2\Lambda_\E^2{\M_\J^2\over F_{\mc R}}
			+{\M_\J^4\over F_{\mc R}^2}\paren{\ln{\M_\J^2/F_{\mc R}\over\Lambda_\E^2}-{1\over2}}
			}
		+\Or{\Lambda_\E^{-2}}.
}
The quartic and quadratic divergences are canceled by the same counter term~\eqref{power ct}.
This time, a natural choice to cancel the logarithmic divergence would be, instead of Eq.~\eqref{log ct},
\al{
V^\tx{c.t.I}_\tx{log}
	&=	{\M_\J^4\over16\pi^2}
			\ln{\mu^2\over\Lambda_\E^2}.
				\label{log ct I}
}
Then the bare potential becomes
\al{
V_\B^\tx{I}
	&=	V_\R^\tx{\red{I}}+V_\tx{power}^\tx{c.t.}+V_\tx{log}^\tx{c.t.I}\nn
	&=	V_\R^\tx{\red{I}} 
		+{1\over16\pi^2}\br{
			\Lambda_\E^4F_{\mc R}^2\sqbr{\ln\fn{F_\Psi^2{F_{\mc R}\Lambda_\E^2\over\mu_0^2}}-{1\over2}}
			+2\Lambda_\E^2F_{\mc R}{\M_\J^2}
			+{\M_\J^4}\ln{\mu^2\over\Lambda_\E^2}
			},
}
and we obtain
\al{
V\fn{\varphi}
	&=	V_\B^\tx{I}+\Delta V_\tx{eff}\nn
	&=	V_\R^\tx{\red{I}}\fn{\varphi,\mu}
		+\red{\Delta V_\R^\tx{\red{I}}\fn{\varphi,\mu},}
		\label{running potential in E}
}
where
\al{
\red{\Delta V_\R^\tx{\red{I}}\fn{\varphi,\mu}}
	:=	-{\sqbr{\M_\J\fn{\varphi}}^4\over16\pi^2}\paren{\ln{\sqbr{\M_\J\fn{\varphi}}^2/F_{\mc R}\over\mu^2}-{1\over2}}.
		\label{correction in I}
}
Again the $\mu$-independence of $V$ fixes the running of the couplings. The running of quartic coupling becomes the same as in Eq.~\eqref{running of lambda} because $F_{\mc R}=1+\mc O\fn{\varphi^2}$ and hence the $\varphi^4$ term is not affected by $\ln F_{\mc R}$; see the discussion below.

When we want to minimize the second term in Eq.~\eqref{running potential in E}, we may choose the renormalization scale
\al{
\mu\sim{\M_\J\over\sqrt{F_{\mc R}}}.
	\label{mu in I}
}
This result reproduces the prescription I in the sense of Eq.~\eqref{prescriptions given}, namely $\mu\sim{\varphi/\sqrt{1+\xi{\varphi^2/\MP^2}}}$ for $F_\Psi=F_\tx{Y}=1$; see also footnote~\ref{relation to ordinary xi}.

\subsection{Where the difference comes from}
Let us summarize the difference between the prescriptions I and II.
The difference of prescriptions I and II comes from that of the subtractions of logarithmic divergence in Eqs.~\eqref{log ct} and \eqref{log ct I}:
\al{
V_\tx{log}^\tx{c.t.I}-V_\tx{log}^\tx{c.t.II}
	=	{\M_\J^4\over16\pi^2}\ln{\Lambda_\J^2\over\Lambda_\E^2}
	=	{\M_\J^4\over16\pi^2}\ln F_{\mc R}
	=	{\xi_{\mc R}y^4\over16\pi^2}{\varphi^6\over M^2}+\cdots,
		\label{difference}
}
where we used Eq.~\eqref{two cutoffs}.
This difference amounts to the finite renormalization of $V_\R$.
Note that the difference~\eqref{difference} is analytic around $\varphi=0$
and that it has only higher order terms with $n\geq6$.

\magenta{
Originally the prescription has been introduced as a choice of frame in which the theory is defined, and it was believed that the radiative correction to the effective potential is minimized by the choice of the renormalization scale as in Eq.~\eqref{prescriptions given}.
As we have shown, however, the physics does not depend on the choice of the frame in which the theory is defined. Instead, for a given renormalized tree-level action, the difference of the prescriptions~\eqref{prescriptions given} can be understood as that of the logarithmic counter terms~\eqref{difference}: The different counter terms lead to the different scales~\eqref{mu in II} and \eqref{mu in I} that minimize the radiative corrections.
}

\subsection{Renormalized potential}


Theoretically, the potential $V_\R$ in Eq.~\eqref{running potential in J} or \eqref{running potential in E},
\al{
V_\R
	&=	\sum_\tx{$n$: even, $n\geq0$}\lambda_{n\R}\fn{\mu}{\varphi^n\over M^{n-4}},
}
may take arbitrary form, so long as it is analytic around $\varphi=0$.
How do we determine its form?

We may reproduce the ordinary Higgs inflation~\cite{Bezrukov:2007ep} that does not assume the criticality, by tuning the infinite number of bare couplings such that $V_\R$ becomes
\al{
\left.V_\R\fn{\varphi,\mu}\right|_{\mu\sim M}
	&\simeq
		\lambda_{4\R}\fn{\mu}\varphi^4,
			\label{f condition}
}
where all the couplings $\lambda_{n\R}\fn{\mu}$ with $n\neq4$ are suppressed at $\mu=M$.
When the form~\eqref{f condition} is put into Eqs.~\eqref{running potential in J} and \eqref{running potential in E}, which result from the counter terms~\eqref{log ct} and \eqref{log ct I}, we obtain the Higgs potential in the prescriptions II and I in the ordinary context, respectively:
\al{
V^\tx{II}
	&=	\lambda_{4\R}\fn{\mu}\varphi^4-{\M_\J^4\over16\pi^2}\paren{\ln{\M_\J^2\over\mu^2}-{1\over2}},
		\label{VJ BS}\\
V^\tx{I}
	&=	\lambda_{4\R}\fn{\mu}\varphi^4-{\M_\J^4\over16\pi^2}\paren{\ln{\M_\J^2/F_{\mc R}\over\mu^2}-{1\over2}}.
		\label{VE BS}
}

However, we may as well obtain the potential of the form of $V^\tx{I}$ in Eq.~\eqref{VE BS} even when we employ the counter term~$V_\tx{log}^\tx{c.t.II}$ in Eq.~\eqref{log ct}
if we choose the following form of the tree-level potential~$V_\R$ in Eq.~\eqref{running potential in J},
\al{
V_{\R}\fn{\varphi,\mu}
	&=	\lambda_{4\R}\fn{\mu}\varphi^4+{\sqbr{\M_\J\fn{\varphi}}^4\over16\pi^2}\ln F_{\mc R}\fn{\varphi},
		\label{f condition another}
}
instead of the form~\eqref{f condition}.
From the same counter term $V_\tx{log}^\tx{c.t.II}$ in Eq.~\eqref{log ct}, we may obtain the forms~\eqref{VJ BS} and \eqref{VE BS} by assuming the tree-level potentials~\eqref{f condition} and \eqref{f condition another}, respectively.
Recall that the second term in the right-hand side of Eq.~\eqref{f condition another} modifies only the higher dimensional terms of $\mc O\fn{\varphi^6}$.
There are infinitely many possibilities for the tree-level potential.
%
%
Therefore there is no reason to suppose one (or any) of the above two as the proper tree-level potential if we restrict ourselves to the low-energy effective field theory.\footnote{
The tree-level potential is determined once we fix the underlying UV-finite theory;
see e.g.\ Refs.~\cite{Hamada:2013mya,Hamada:2015ria} for discussions based on string theory.
}

\section{Flattening Higgs potential by gauge kinetic function}\label{gauge kinetic}
Let us turn to more realistic running of the quartic coupling in the SM:
\al{
V_\tx{SM}
	&=	\lambda_{4\R}\fn{\mu}\varphi^4+\red{\Delta V_\R\fn{\varphi,\mu}},
}
where \red{$\Delta V_\R$} is the \red{finite correction~\eqref{finite correction II} or \eqref{correction in I} in the counterterm formalism;} see also footnote~\ref{footnote on lambda}. 
In the SM, $\beta_{4\R}:={\df\over\df\ln\mu}\lambda_{4\R}$ turns from negative to positive around the scale $\mu_\tx{min}\sim10^{17}\,\tx{GeV}$, and we may approximate as~\cite{Hamada:2014iga,Hamada:2014wna}
\al{
\lambda_{4\R}\fn{\mu}
	&\approx	\lambda_{4\R}^\tx{min}+b_{4\R}\,\paren{\ln{\mu\over\mu_\tx{min}}}^2,
		\label{lambda approximated}
}
where $b_{4\R}$ can be computed within the SM as
\al{
b_{4\R}
	&\simeq	{0.1\over\paren{16\pi^2}^2}
	\simeq	5\times10^{-6}.
		\label{value of b}
}
The negative $\beta_{4\R}$ for $\mu<\mu_\tx{min}$ is dominated by top quark loop, while the positive $\beta_{4\R}$ for $\mu>\mu_\tx{min}$ by the $U(1)_Y$ and $SU(2)_L$ gauge boson loops.

For top quark loop, the contribution is through the effective mass~$\M_\J=y\varphi F_\tx{Y}/F_\Psi$. In the prescription I in the sense of Eq.~\eqref{prescriptions given}, namely in Eq.~\eqref{VE BS} with the tree-level potential~\eqref{f condition}, we get the constant $\mu$ in the large $\varphi$ limit,
\al{
\mu
	\sim {\M_\J\over\sqrt{F_{\mc R}}}
	\to {\MP\over\sqrt{\xi}},
	\label{nice behavior}
}
and the effective quartic coupling $\left.\lambda_{4\R}\fn{\mu}\right|_{\mu\sim\M_\J/\sqrt{F_{\mc R}}}$ stops running for large $\varphi$~\cite{Allison:2013uaa}.\footnote{
Here we have assumed $F_\tx{Y}=F_\Psi=1$. If $F_\tx{Y}=1+\xi_\tx{Y}\varphi^2/M^2$ and $F_\Psi=1+\xi_\Psi\varphi^2/M^2$, we obtain
\als{
\mu\sim{\M_\J\over\sqrt{F_{\mc R}}}\to {\xi_\tx{Y}\over\sqrt{\xi}\xi_\Psi}\MP
}
instead of Eq.~\eqref{nice behavior}; see also footnote~\ref{relation to ordinary xi}.
}
This mechanism makes the potential even flatter at the SM criticality and helps to earn a sufficiently large $e$-folding number for smaller $\xi\sim10$;
in the prescription II we lack this mechanism and need larger $\xi\sim10^2$~\cite{Hamada:2014iga,Hamada:2014wna}.

Similarly, 
the contribution of the gauge boson loop is through~\cite{Hamada:2014wna}
\al{
\M^\tx{gauge}_\J
	&=	g\varphi\sqrt{F_\Phi\over F_\tx{g}},
}
where $g$ is the gauge coupling and $F_\tx{g}$ is the gauge kinetic function, namely the function of $\varphi$ in front of the gauge kinetic term.

When we raise the scale beyond $\mu>\mu_\tx{min}$ in the SM, the top Yukawa coupling becomes smaller and smaller.
To the first approximation, the running at $\mu>\mu_\tx{min}$ is governed by the gauge boson loop. Then in the prescription~II in the ordinary context, which corresponds to Eq.~\eqref{VJ BS} with the tree-level potential~\eqref{f condition}, the effective potential becomes
\al{
V_\tx{SM}=\left.\lambda_{4\R}\fn{\mu}\varphi^4\right|_{\mu=\M_\J^\tx{gauge}}.
}
When we assume that $F_\Phi\simeq1$, we obtain
\al{
\M^\tx{gauge}_\J
	&=	{g\varphi\over\sqrt{1+\xi_\tx{g}{\varphi^2\over M^2}}}.
		\label{gauge mass}
}
\magenta{In the large-$\varphi$ limit,}
\al{
\M^\tx{gauge}_\J
	\to	{g\over\sqrt{\xi_\tx{g}}}M.
}
We propose that this can be used in the prescription~II in the ordinary context as an alternative mechanism to Eq.~\eqref{nice behavior} in order to stop the running of quartic coupling $\left.\lambda_{4\R}\fn{\mu}\right|_{\mu\sim\M_\J^\tx{gauge}}$ for large $\varphi$.\footnote{
When the top Yukawa contribution is non-negligible, one may further introduce e.g.\
\als{
F_\Psi\fn{\varphi}
	&=	\sqrt{1+2\xi_\Psi{\varphi^2\over M^2}}
	=	1+\xi_\Psi{\varphi^2\over M^2}-{\xi_\Psi^2\over2}{\varphi^4\over M^4}+{\xi_\Psi^3\over2}{\varphi^6\over M^6}+\cdots
		\label{F_Psi non minimal}
}
together with $F_\tx{Y}=1$, which stops running due to the top contribution too:
\als{
\M_\J
	&=	{y\varphi\over\sqrt{1+2\xi_\Psi{\varphi^2\over M^2}}}
	\to	{y\over\sqrt{2\xi_\Psi}}M.
}
}

\magenta{
The $\varphi$-dependent mass~\eqref{gauge mass} takes the same form as the prescription I in Eq.~\eqref{prescriptions given} if we neglect the gauge coupling $g$. Therefore, for example, we may set $\xi_\tx{g}=\xi_{\mc R}$ and $M=\MP$, then the subsequent analysis becomes identical to those in Ref.~\cite{Hamada:2014wna}. This serves as an explicit example of viable parameter set that realizes the above-mentioned idea. 
}

\section{Summary}\label{summary}
We have analyzed the one-loop effective action in the simplified Higgs-Yukawa model, which captures essential features of the Higgs potential in the Higgs inflation.
We have shown that the effective actions obtained in the Jordan and Einstein frames are exactly the same if we properly take into account the change of path integral measure.
We show that, in the counterterm formalism, the prescriptions I and II are merely two specific choices of counter terms to cancel the logarithmic divergence.
We point out that the difference between I and II can be absorbed into the choice of tree-level potential, including higher dimensional terms, from infinitely many possibilities.


We have also proposed a mechanism to stop the running of the effective quartic coupling in the prescription~II in the ordinary context, using the gauge kinetic function: $\mu\sim \varphi/\sqrt{F_\tx{g}}\to M/\sqrt{\xi_\tx{g}}$.
Detailed phenomenological study of this scenario will be presented in a separate publication.

We briefly comment on the remaining points to be addressed: In this paper, we have concentrated on the fermion loop. It would be worth including the scalar loop, as in Ref.~\cite{Kamenshchik:2014waa}, and also the gauge boson loop. It is also worth studying the issue of gauge dependence in these loops.

\subsection*{Acknowledgment}
We thank Seong Chan Park, Mikhail Shaposhnikov, and Satoshi Yamaguchi for useful discussions. The work of Y.H.\ and K.O.\ are supported in part by JSPS KAKENHI Grant Nos.~16J06151 and 23104009, 15K05053, respectively.

\appendix
\section*{Appendix}
\section{Classical dynamics in terms of Jordan-frame field}\label{mechanism}
When we consider the classical dynamics of scalar field under gravity, it is convenient to define the canonically normalized scalar field
\al{
\df\chi
	&=	\G\fn{\varphi}\,\df\varphi,
}
where
\al{
\G\fn{\varphi}
	&:=	\sqrt{
			{F_\Phi\fn{\varphi}\over F_{\mc R}\fn{\varphi}}
			+{3\over2}\paren{\MP F_{\mc R}'\fn{\varphi}\over F_{\mc R}\fn{\varphi}}^2
			}.
}
We use the Friedmann-Lema\^itre-Robertson-Walker ansatz\footnote{
We neglect the spatial curvature, but may recover it by ${\df\bs x}^2\to{\df\bs x}^2+K{\paren{\bs x\cdot\df\bs x}^2\over1-K\bs x^2}$.
}
\al{
g_{\mu\nu}^\E\df x^\mu\df x^\nu
	&=	-{\df t_\E}^2+\paren{a_\E\fn{t_\E}}^2{\df\bs x}^2.
}
In the Einstein frame, the Einstein equations reduce to the ordinary Friedmann equation:\footnote{
We may recover the spatial curvature by $H_\E^2\to H_\E^2+{K\over a^2}$.
}
\al{
H_\E^2
	&=	{\rho_\E\over3\MP^2},\\
{\df\rho_\E\over\df t_\tx{E}}
	&=	-3\paren{\rho_\E+p_\E}H_\E,
}
where
\al{
\rho_\E
	&=	{1\over2}\paren{\df\chi\over\df t_\E}^2+U_\E\fn{\varphi},\\
p_\E
	&=	{1\over2}\paren{\df\chi\over\df t_\E}^2-U_\E\fn{\varphi},
}
and $H_\E:={1\over a_\E}{\df a_\E\over\df t_\E}$.

The Higgs-field equation reads
\al{
{\df^2\chi\over\df t_\E^2}+3H_\E{\df\chi\over\df t_\E}
	&=	-{\df U_\E\over\df \chi}.
}
In terms of the Jordan-frame field,
\al{
{\df^2\varphi\over\df t_\E^2}+{\df\varphi\over\df t_\E}\paren{3H_\E+{\df\over\df t_\E}\ln \G}
	&=	-{1\over\G^2}{\df U_\E\over\df\varphi}.
}
The universe expands with the rate $H_\E$, whereas the Jordan-frame field~$\varphi$ receives extra friction ${\df\over\df t_\E}\ln \G$, and rolls slower (faster) than under its absence when it is positive (negative).
This term will turn out to be the same order as the slow-roll parameter under the slow-roll condition shown below.

We assume the slow-roll inflation. The slow-roll parameters read
\al{
\epsilon
	&:=	{\MP^2\over2U_\E^2}\paren{\df U_\E\over\df\chi}^2
	=	{\MP^2\over 2U_\E^2\G^2}\paren{\df U_\E\over\df\varphi}^2
	\ll1,
		\label{epsilon_V defined}\\
\eta
	&:=	{\MP^2\over U_\E}{\df^2 U_\E\over\df \chi^2}
	=	{\MP^2\over U_\E\G}{\df\over\df\varphi}\paren{{1\over\G}{\df U_\E\over\df\varphi}}
	\ll1.
}
The Friedmann and Higgs-field equations become, respectively,
\al{
3\MP^2H_\E^2
	&=	U_\E,\\
3H_\E{\df\chi\over\df t_\E}
	&=	-{\df U_\E\over\df\chi}.
}
In terms of the Jordan-frame field, the latter reads
\al{
3H_\E{\df\varphi\over\df t_\E}
	&=	-{1\over \G^2}{\df U_\E\over\df\varphi}
	=:	-{\df\mc U_\E\over\df\varphi},
}
where we have defined the effectual potential
\al{
\mc U_\E
	&=	\int\df\varphi {1\over\mc G^2}{\df U_\E\over\df\varphi}+\tx{const.},
		\label{effectual potential}
}
which takes into account the effect from $\mc G$.
Using this potential, the slow-roll parameters can be rewritten as
\al{
\epsilon
	&=	{\mc G^2\MP^2\over2U_\E^2}\paren{\df\mc U_\E\over\df \varphi}^2,\\
\eta
	&=	{\MP^2\over U_\E}\paren{{\df^2\mc U_\E\over\df\varphi^2}+{\df\mc U_\E\over\df\varphi}{\df\over\df\varphi}\ln\mc G},
}
though this expression may not be particularly convenient.
In this paper, we omit the subscript $\E$ from $U_\E$ and $\mc U_\E$ which are always given in the Einstein frame.

\section{Examples that lead to various Higgs inflations}\label{various}

Even if we decide to take the simple form of the tree-level potential~\eqref{f condition} in Eq.~\eqref{running potential in J} or \eqref{running potential in E},
we still have freedom to choose any form of $F_X\fn{\varphi}$s.
We review several examples that lead to viable cosmic inflations:
\begin{itemize}
\item In Ref.~\cite{Hamada:2014wna}, the authors have spelled out the result from prescriptions I and II in the ordinary context, with the tree-level potential~\eqref{f condition} in Eq.~\eqref{running potential in E} and \eqref{running potential in J}
, respectively, and with the function $F_X=1$ except for $F_{\mc R}=1+\xi_{\mc R}{\varphi^2\over M^2}$. The former prescription~I allows smaller $\xi:=\xi_{\mc R}\MP^2/M^2\sim10$ because the coupling stops running for $\varphi\gg M/\sqrt{\xi_{\mc R}}$:
\al{
\mu
	\sim {\varphi\over\sqrt{F_{\mc R}}}
	\to {M\over\sqrt{\xi_{\mc R}}}.
	\label{freezing in I}
}
The latter prescription~II can have a chaotic inflation for $\xi\sim10^2$, since the effectual potential~\eqref{effectual potential} becomes
\al{
\mc U
	&\sim
		\tx{const.}+{\beta_{\lambda}\MP^2\over48\xi^2}\varphi^2,
}
due to
\al{
\G
	&\to {\sqrt{6}\MP\over\varphi}
}
for large $\xi_{\mc R}$; see Appendix~\ref{mechanism}.
\item When we have large $\xi_\Phi$ only, in particular with $F_{\mc R}=1$ which gives $U=V={\lambda\over4}\varphi^4$, we get
\al{
\G\to\sqrt{\xi_\Phi}\,{\varphi\over M}
}
and hence
\al{
\mc U
	&=	\tx{const.}+{\lambda M^2\over2\xi_\Phi}\varphi^2.
}
This can also lead to a chaotic inflation when $\lambda/\xi_\Phi\ll1$~\cite{Nakayama:2014koa}.
\item When we terminate the tree-level potential at the $2n$th order
\al{
V=\sum_\tx{$n'$: even, $4\leq n'\leq2n$}\lambda_{n'}{\varphi^{n'}\over M^{n'-4}},
}
instead of Eq.~\eqref{f condition}, and the function at the $n$th order
\al{
F_{\mc R}=\sum_\tx{$n'$: even, $0\leq n'\leq n$}\xi_{\mc R,n'}{\varphi^{n'}\over M^{n'}},
}
the resultant classical potential becomes constant~\cite{Park:2008hz}:
\al{
U\to{\lambda_{2n}\over\xi_{\mc R,n}^2}M^4.
}
\end{itemize}

\bibliographystyle{TitleAndArxiv}
\bibliography{Bibliography}

\end{document}